\documentclass[aps,prl,twocolumn,superscriptaddress,nopacs]{revtex4}
\usepackage{graphicx}

\usepackage{dcolumn}

\usepackage{bm}
\usepackage{color}
\usepackage[colorlinks]{hyperref}
\input{epsf}
\begin{document}

%\preprint{}

\title{Photo-excited zero-resistance states in the GaAs/AlGaAs system}

\author{R. G. Mani}
\email{mani@deas.harvard.edu} \affiliation {Harvard University,
Gordon McKay Laboratory of Applied Science, 9 Oxford Street,
Cambridge, MA 02138, USA}

%\author{}
%\thanks{}
%\affiliation{}
\date{\today}
\begin{abstract}
The microwave-excited high mobility two-dimensional electron
system exhibits, at liquid helium temperatures, vanishing
resistance in the vicinity of $B = [4/(4j+1)] B_{f}$, where $B_{f}
= 2\pi\textit{f}m^{*}/e$, m$^{*}$ is an effective mass, e is the
charge, and \textit{f} is the microwave frequency. Here, we
summarize some experimental results.

Journal-Reference: International Journal of Modern Physics B
\textbf{18}, 3473 (2004)
\end{abstract}

%\pacs{ Journal-Ref: Phys. Rev. B \textbf{69}, 193304 (2004)}

\maketitle

%%%%%%%%%%%%%%%%%%%%%%%%%%%%%%%%%%%%%%%%%%%%%%%%%%%%%%%%%%%%
% The main text of your paper   begins here              %
%%%%%%%%%%%%%%%%%%%%%%%%%%%%%%%%%%%%%%%%%%%%%%%%%%%%%%%%%%%%

\section{Introduction}
The creation, detection, and study of exciton condensates in
semiconductors constitute interesting problems in condensed matter
physics.\cite{1} Exciton populations are usually created by
photo-exciting a semiconductor, so that the electron-hole pairs
consist of conduction band electrons and valence band holes. Then,
optical techniques are employed to detect and study the excitonic
system.\cite{1,2} Recently, there has also been interest in the
possibility of \textit{electrical detection}, and the realization
\textit{without photoexcitation}, of such exotic states in a
bilayer quantum Hall system, where the excitons result from the
pairing of electrons in one quantum Hall layer with holes in the
neighboring layer.\cite{1,3,4} This report examines a hybrid
scenario where \textit{photoexcitation} serves to create
inter-Landau level, instead of inter-band, excitations in a
quantum Hall system, as \textit{electrical detection} is employed
to detect the system response.\cite{5,6,7,8,9,10,11,12,13} In
particular, we study \textit{transport} in a semiconductor
single-layer two-dimensional electron system, where
\textit{microwave photo-excitation} produces a steady state
density of inter-Landau level excitations ("excitons").
Surprisingly, such a system exhibits novel zero-resistance states
without concomitant Hall quantization, at low temperatures, $T$,
about magnetic fields, $B$ = $(4/5) B_{f}$ and $B$ = $(4/9)
B_{f}$, where $B_{f}$ = $2\pi f m^{*}/e$, $m^{*}$ is an effective
mass, $e$ is the electron charge, and $f$ is the radiation
frequency.\cite{9} The experimental observations have led to a
broad theoretical study of this
situation.\cite{14,15,16,17,18,19,20,21,22,23,24,25,26,27,28,29,30,31,32,33,34,35}

At the present, it is uncertain whether the observed
zero-resistance states should be viewed as a novel condensate with
a role for excitons, although we had suggested this
possibility.\cite{9,32} Theoretical studies have realized
radiation-induced magneto-resistivity oscillations through various
approaches.\cite{14,16,18,22,23,24,28,29,30,31} The observed
zero-resistance states have then been attributed to a negative
resistivity instability, and current-domain formation.\cite{17} As
there remain many open questions from both the experimental and
theoretical perspectives, we  examine here just the experimental
situation, and refer the reader to the literature for further
theory.

\section{Experiment}
%%%%%%%%%%%%%%%%%%%%%%%%%%%%%%%%%%%%%%%%%%%%%%%%%%%%%%%%%%%%
%                                                          %
% You may repeat \section{SECTION N-th HEADING TYPE HERE}  %
%                                                          %
% Do start a subsection or sub-subsection, do this:-       %
%                                                          %
%   \subsection{SUBSECTION HEADING TYPE HERE}              %
%                                                          %
%   \subsubsection{SUB-SUBSECTION HEADING TYPE HERE}       %
%                                                          %
% instead of the above                                     %
%                                                          %
%%%%%%%%%%%%%%%%%%%%%%%%%%%%%%%%%%%%%%%%%%%%%%%%%%%%%%%%%%%%

\begin{figure}[b]
%h=here, t=top, b=bottom, p=separate figure page
\begin{center}\leavevmode
\includegraphics[scale = 0.25,angle=0,keepaspectratio=true,width=3.35in]{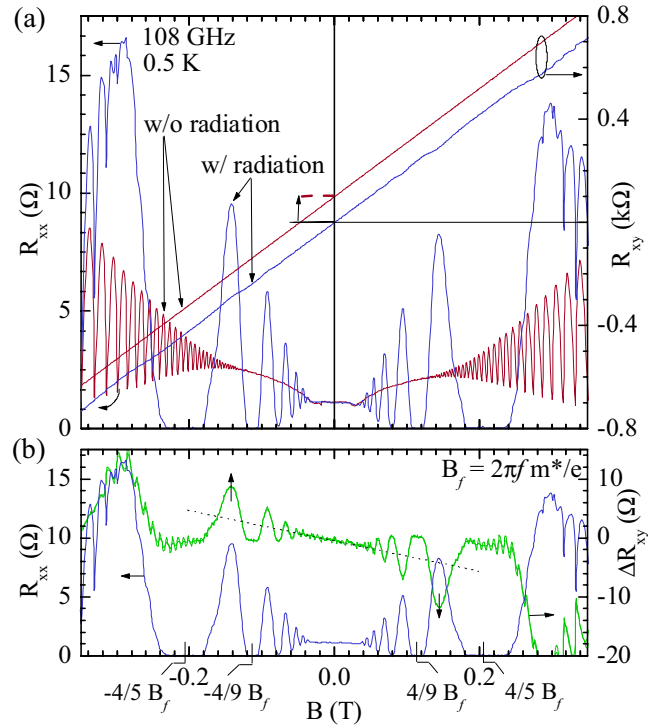}
\caption{ (a): Transport measurements with (w/) and without (w/o)
microwave radiation at $f$ = 108 GHz. Radiation induced vanishing
$R_{xx}$ is observable about $(4/5) B_{f}$ and $(4/9) B_{f}$. A
comparison of the w/ and w/o radiation $R_{xy}$ indicates
antisymmetric-in-B oscillations in $R_{xy}$ under photoexcitation,
which correlate with the $R_{xx}$ oscillations. Here, the w/o
radiation Hall data have been offset for the sake of clarity. (b):
The radiation induced change in the Hall resistance, $\Delta
R_{xy}$, is shown along with $R_{xx}$. Note that the
antisymmetric-in-B $\Delta R_{xy}$ oscillations.} \label{figF}
\end{center}
\end{figure}

Experiments were carried out on standard devices fabricated from
GaAs/AlGaAs heterostructures. The best material was typically
characterized by an electron density, $n$(4.2 K) $\approx$ $3$
$\times$ $10^{11}$ cm$^{-2}$, and an electron mobility, $\mu$, up
to $1.5$ $\times$ $10^{7}$ cm$^{2}$/Vs. Standard  four-terminal
lock-in based electrical measurements were performed with the
sample mounted inside a waveguide and immersed in pumped liquid
He-3 or He-4. Electromagnetic (EM) waves in the microwave part of
the spectrum were generated using tunable sources. The EM waves
were directed to the sample through the waveguide. Further details
appear elsewhere.\cite{9}

\begin{figure}[t]
%h=here, t=top, b=bottom, p=separate figure page
\begin{center}\leavevmode
\includegraphics[scale = 0.25,angle=0,keepaspectratio=true,width=3.35in]{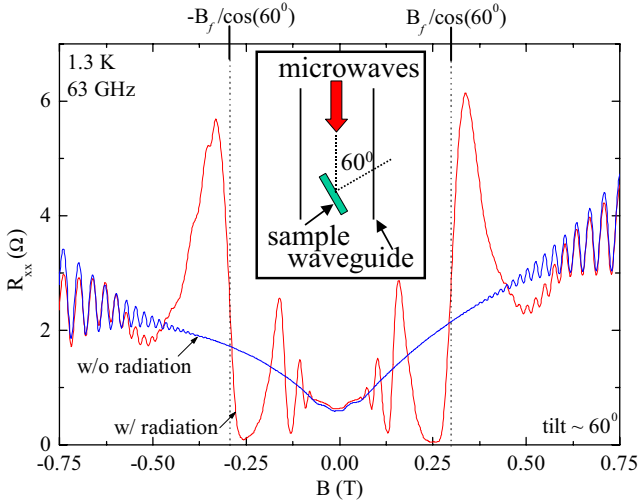}
\caption{The effect of microwave radiation on a tilted 2DES, when
the tilt angle is 60$^{0}$, see inset. Here, the abscissa shows
the applied (total) magnetic field. Typically, the
radiation-induced magnetoresistance oscillations are mostly
sensitive to the normal component of $B$. Thus, the characteristic
$B$-scale under tilt is given by $B_{f}/\cos (\theta)$, where
$\theta$ is the tilt angle. } \label{figN}\end{center}\end{figure}

The data of Fig. 1(a) illustrate the radiation-induced effect on
the diagonal ($R_{xx}$) and Hall resistances ($R_{xy}$), when the
specimen is excited with microwaves at $f$ = 108 GHz. In Fig.
1(a), the application of microwaves induces giant oscillations in
$R_{xx}$, which are characterized by the property that the
$R_{xx}$ under photoexcitation drops below the dark $R_{xx}$ over
wide $B$ intervals. Thus, at approximately $B_{f}/B$ = $j$ and
$B_{f}/B$ = $j+1/2$, with $j$ = 1,2,3... , the $R_{xx}$ under
photoexcitation is equal to the dark $R_{xx}$. For $j < B_{f}/B <j
+ 1/2$, the photo-excited $R_{xx}$ is reduced with respect to the
dark $R_{xx}$. On the other hand, for $j+1/2 < B_{f}/B < j+1$, the
irradiated $R_{xx}$ is enhanced with respect to the dark $R_{xx}$.
A remarkable feature in the data are the zero-resistance states
which are manifested about $B_{f}/B = j + 1/4$, i.e., in the
vicinity of $B = (4/5) B_{f}$ and $B = (4/9)B_{f}$. These $R_{xx}$
minima exhibit activated transport vs. the temperature, which is
suggestive, from the experimental perspective, of a possible
radiation-induced gap in the electronic spectrum.\cite{9}

The data of Fig. 1a also illustrate the influence of radiation on
the Hall effect. Here, a comparison of the dark $R_{xy}$ with the
photo-excited $R_{xy}$ reveals subtle microwave induced
oscillations, which are period commensurate with the radiation
induced $R_{xx}$ oscillations.\cite{9,12} Yet, there do not occur
microwave-induced plateaus in the Hall effect that coincide with
the radiation-induced vanishing $R_{xx}$ in the vicinity of B =
$(4/5) B_{f}$ and $(4/9) B_{f}$. In order to highlight these
radiation-induced changes in $R_{xy}$, $\Delta R_{xy} =
R_{xy}^{excited} - R_{xy}^{dark}$ is shown along with $R_{xx}$ in
Fig. 1(b). A comparison of Fig. 1(a) and Fig. 1(b) shows clearly
that $R_{xy}$ is reduced in magnitude over the B-intervals, where
$R_{xx}$ is enhanced by the radiation. On the other hand, as the
diagonal resistance vanishes upon microwave excitation, as in the
vicinity of $B = (4/5) B_{f}$, for example, the correction $\Delta
R_{xy}$ also vanishes.
\begin{figure}[t]
%h=here, t=top, b=bottom, p=separate figure page
\begin{center}\leavevmode
\includegraphics[scale = 0.25,angle=0,keepaspectratio=true,width=3.35in]{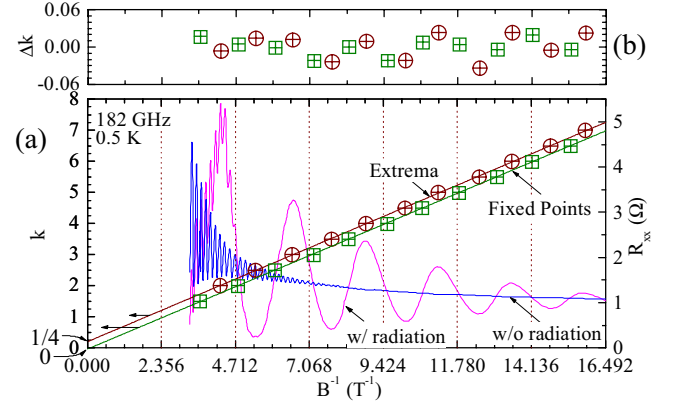}
\caption{(a) Right ordinate: $R_{xx}$ with (w/) and without (w/o)
radiation at 182 GHz is plotted vs. $B^{-1}$. Left ordinate:
Half-cycle plots of the extremal- and fixed- points in the
radiation induced resistance oscillations. The slope of the linear
fit is proportional to $m^{*}/m$, while the ordinate intercepts
measure the phase, which are "1/4" and "0" for the extremal and
fixed points, respectively. (b): The residue in the labels,
$\Delta k = k - k_{FIT}$, is shown as a function of $B^{-1}$.}
\label{figL}\end{center}\end{figure}

\begin{figure}[t]
%h=here, t=top, b=bottom, p=separate figure page
\begin{center}\leavevmode
\includegraphics[scale = 0.25,angle=0,keepaspectratio=true,width=3.35in]{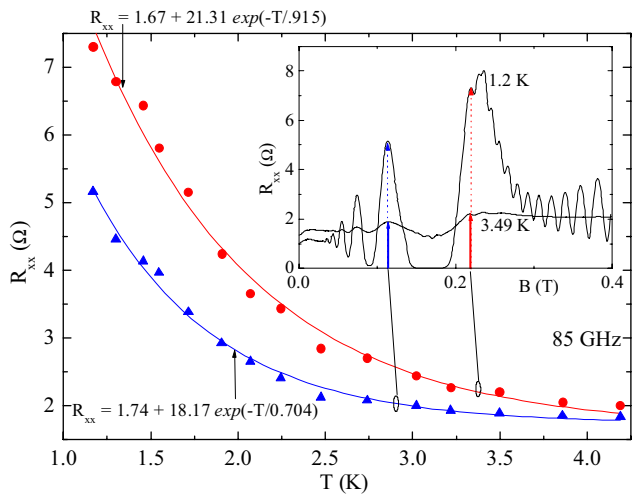}
\caption{(a) Inset: The diagonal resistance, $R_{xx}$, is
exhibited vs. the magnetic field, $B$, with microwave excitation
at 85 GHz. The main panel exhibits the temperature dependence of
the first and second resistance maxima, along with empirical fits
to the data.} \label{figD}
\end{center}
\end{figure}

As the period of the radiation induced oscillations (see Fig. 1)
are mostly sensitive to the $B$-component that is perpendicular to
the plane of the 2DES, they tend to span a wider $B$-interval, at
a finite tilt angle of the specimen. Thus, under tilt, the
characteristic field scale becomes $B_{f}/cos(\theta)$, and the
dark $R_{xx}$ intercepts the photo-excited $R_{xx}$ in the
vicinity of $B=[B_{f}/cos(\theta)]/j$ and
$B=[B_{f}/cos(\theta)]/(j+1/2)$, with $j$ = 1,2,3... These
features can be detected in Fig. 2

\begin{figure}[t]
%h=here, t=top, b=bottom, p=separate figure page
\begin{center}\leavevmode
\includegraphics[scale = 0.25,angle=0,keepaspectratio=true,width=3.35in]{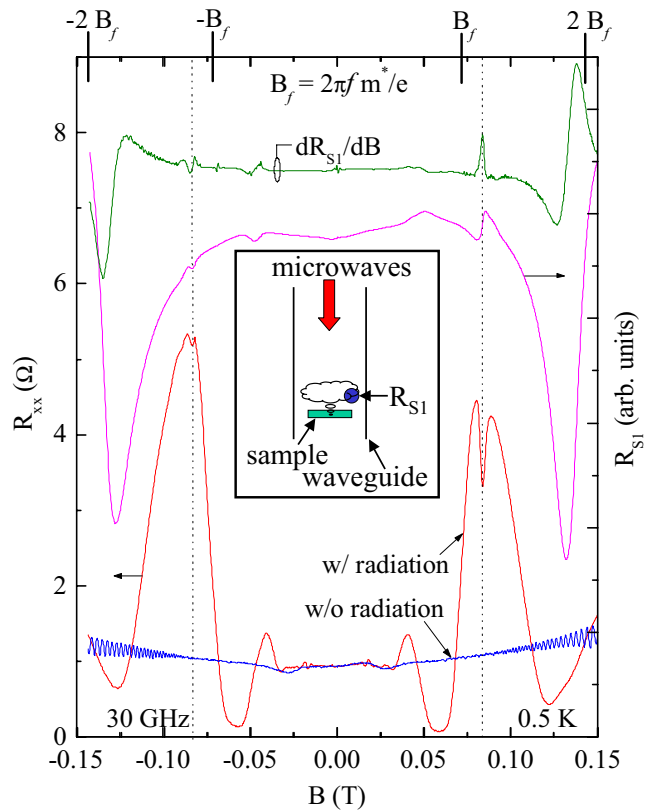}
\caption{This figure examines the reflection/emission
characteristics of the irradiated 2DES. Thus, the figure exhibits
the dark, and the 30 GHz photo-excited $R_{xx}$, as well as the
sensor $R_{S1}$ response under photoexcitation. Also shown is the
numerical derivative, $dR_{S1}/dB$. The $R_{S1}$ signal shows a
pair of significant features between $B_{f}$ and 2$B_{f}$, as the
near 2$B_{f}$ feature coincides with a second harmonic, two-photon
process. The inset illustrates the measurement configuration.}
\label{figN}\end{center}\end{figure}

The phase of the radiation induced $R_{xx}$ oscillations is an
important parameter because it can be used to compare experiment
to theory.\cite{9,10,14,16,18} To determine this observable,
$R_{xx}$ maxima and minima of 182 GHz data (see Fig. 3) are
initially labelled by integers and half-integers, respectively,
assuming a cosine waveform. The extrema labels, denoted by the
index $k$ in Fig. 3, are then plotted vs. extremal values of
$B^{-1}$, and a linear fit, $k_{FIT}$, is applied to determine the
slope, $dk_{FIT}/dB^{-1}$ = $B_{f}$, and the ordinate intercept.
In Fig. 3, the ordinate intercept for the fit of extremal data
points at 1/4 demonstrates a 1/4-cycle phase shift of the extrema
with respect to the assumed (cosine) waveform.\cite{9} Thus, the
plot (Fig. 3(a)) demonstrates that the phase is broadly consistent
with a lineshape $R_{xx} \approx -sin (B_{f}/B)$, i.e., resistance
minima about $hf/\hbar\omega_{c} = j + 1/4$, and resistance maxima
about $hf/\hbar\omega_{c} = j + 3/4$. A similar analysis of the
nodes or fixed points suggests that they are characterized by
$hf/\hbar\omega_{c} = j$, or $hf/\hbar\omega_{c} = j + 1/2$, see
Fig. 3.
\begin{figure}[t]
%h=here, t=top, b=bottom, p=separate figure page
\begin{center}\leavevmode
\includegraphics[scale = 0.25,angle=0,keepaspectratio=true,width=3.35in]{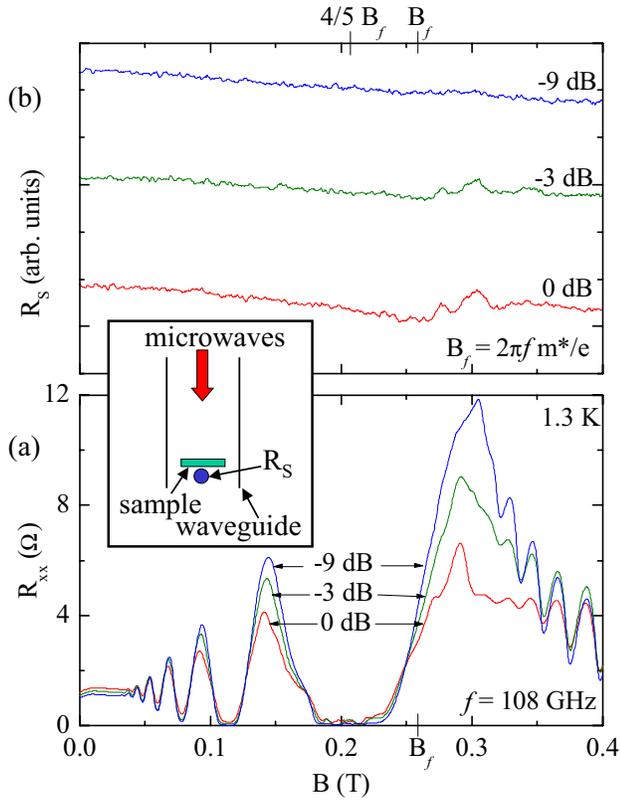}
\caption{The figure illustrates the transmission characteristics
of the 2DES under irradiation. As shown in the inset, a resistance
sensor below the sample serves as the radiation detector. $R_{xx}$
is shown in (a) while the detector resistance, $R_{S}$, is shown
in (b). The bottom panel (a) shows that, when the power
attenuation factor exceeds -9dB, the oscillation amplitude
decreases with increasing radiation intensity. At the same time,
the detector response (b) suggests non monotonic transmission
above $B_{f}$. Oscillations in $R_{xx}$ below $B_{f}$ appear
imperceptible in $R_{S}$.} \label{figP}\end{center}\end{figure}

In prior work, we have shown that the radiation induced resistance
\textit{minima} exhibit activated transport, i.e., $R_{xx} \propto
\exp(-T_{0}/T)$.\cite{9} As a supplement to that previous report,
we examine the $T$-dependence of the resistance \textit{maxima} in
Fig. 4. Thus, the inset in Fig. 4 shows $R_{xx}$ under microwave
excitation, with a pair of marked maxima, whose $T$-evolution is
followed in the main part of the figure. Fig. 4 shows that the
$R_{xx}$ maxima increase rapidly with decreasing temperature, and
the variation can be empirically represented by an exponential
function. The $T$-variation of the maxima have been attributed by
Dmitriev et al., to the $T$-dependence of the inelastic scattering
rate.\cite{31}

Measurements examining possible reflection/emission and
transmission characteristics of the high mobility 2DES under
microwave excitation are illustrated in Fig. 5 and Fig. 6. For
these experiments [see inset, Figs. 5 and 6], a resistance sensor
was placed immediately above or below the sample.\cite{9}

Fig. 5 exhibits the dark, and irradiated at 30 GHz, $R_{xx}$,
along with the sensor $R_{S1}$ response under photoexcitation. The
irradiated $R_{xx}$ shows giant microwave-induced oscillations.
Note that, in this low-$f$ case, the irradiated $R_{xx}$ crosses
the dark $R_{xx}$ also near $2B_{f}$. Also shown in the figure is
$dR_{S1}/dB$, which serves to bring out weak features in the
$R_{S1}$. The $R_{S1}$ signal shows a pair of significant features
between $B_{f}$ and 2$B_{f}$. Here, the largest feature might
originate from a two-photon process. In the derivative signal,
there is a very weak feature near -$B_{f}$ (but not at +$B_{f}$),
which could signify electrically detected cyclotron resonance. The
origin of other structure in the sensor signal is not fully
understood.

Some transmission measurements are shown in Fig. 6. Here, the
optimal radiation induced $R_{xx}$ response was observed in the
vicinity of -9 dB. A further increase of the radiation intensity
(dB $\rightarrow$ 0) produced a reduction in the peak height along
with an increase in the resistance at the minima (see Fig. 6(a)).
At the same time, the transmission sensor (see Fig. 6(b))
indicated structure at $B$ $>$ $B_{f}$, which correlated with a
strong radiation-induced decrease in $R_{xx}$ just above 0.3
Tesla. Observed oscillations in $R_{xx}$ below $B_{f}$ (Fig. 6(a))
were imperceptible, however, in the detector response (see Fig.
6(b)).
\begin{figure}[t]
%h=here, t=top, b=bottom, p=separate figure page
\begin{center}\leavevmode
\includegraphics[scale = 0.25,angle=0,keepaspectratio=true,width=3.35in]{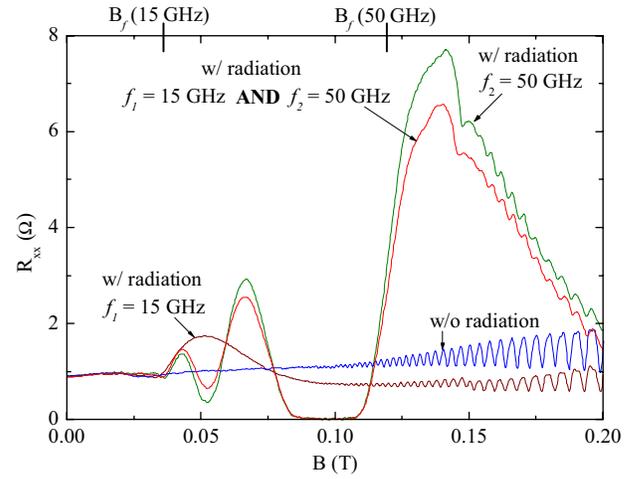}
\caption{This figure traces the effect of simultaneously exciting
the GaAs/AlGaAs device at two microwave frequencies. Thus, we have
exhibited the dark (w/o radiation) magnetoresistance, the $R_{xx}$
obtained under excitation at 50 GHz only, the $ R_{xx}$ with
excitation at 15 GHz only, and the $R_{xx}$ obtained under
simultaneous excitation at 15 GHz and 50 GHz. The latter looks
mostly like the trace obtained at 50 GHz. The characteristic
$B_{f}$ have been marked the figure.}
\label{figM}\end{center}\end{figure}

It is also interesting to examine the effect of simultaneous
excitation of the specimen at two different microwave
frequencies.\cite{15,24} Thus, Fig. 7 exhibits  measurements with
excitation of the specimen at 15 GHz only, and at 50 GHz only.
Here, at 50 GHz, there occur a number of radiation-induced
oscillations and one broad zero-resistance state about $(4/5)
B_{f}$. On the other hand, at 15 GHz, the radiation seems to
produce a non-monotonic variation in $R_{xx}$, mostly above
$B_{f}$. Simultaneous excitation at two-frequencies produces a
$R_{xx}$ trace that looks very much like the trace at the higher
$f$, i.e., 50 GHz. An interesting feature is observed here by
comparing the $f_{1}$ AND $f_{2}$ signals at $B$ $\approx$ 0.07 T
and $B$ $\approx$ 0.14 T. At both these $B$, the peak amplitude of
the $f_{1}$ AND $f_{2}$ signal is reduced with respect to the
$f_{2}$ signal. Yet, one observes that in the $B$ $\approx$ 0.07 T
case, only $f_{1}$ produces an enhancement, while in the $B$
$\approx$ 0.14 T case, it produces a resistance suppression, with
respect to the dark signal. Thus, perhaps, one should attribute
the small decrease in the amplitude of the $f_{1}$ AND $f_{2}$
signal to a "breakdown" type heating effect, due to the increased
radiation intensity resulting from the combined excitation.

\section{Summary}

In summary, radiation induced zero-resistance states and
associated oscillations in the high mobility 2DES exhibit a rich
phenomenology. Here, we have shown that (a) $R_{xx}$ exhibits
saturation at zero-resistance under microwave excitation in the
vicinity of $[4/(4j+1)] B_{f}$, with concomitant vanishing Hall
resistance correction, $\Delta R_{xy}$, see Fig. 1, (b) the period
of the radiation induced oscillations are mostly sensitive of the
normal component of the magnetic field, see Fig. 2, (c) the
oscillatory minima are shifted by 1/4 cycle with respect to the
$hf/\hbar \omega_{c}$ = $j$ condition, see Fig. 3, (d) the
oscillatory resistance maxima increase strongly with decreasing
temperature, see Fig. 4, (e) a response is observable in the
transmission and reflection signals, mostly above $B_{f}$, see
Figs. 5 and 6, and (e) simultaneous photoexcitation at two
radiation frequencies seems not to produce simple superposition,
see Fig. 7.
%%%%%%%%%%%%%%%%%%%%%%%%%%%%%%%%%%%%%%%%%%%%%%%%%%%%%%%%%%%%
% Doing Acknowledgement                          %
%%%%%%%%%%%%%%%%%%%%%%%%%%%%%%%%%%%%%%%%%%%%%%%%%%%%%%%%%%%%

\section*{Acknowledgements}

We acknowledge discussions with K. von Klitzing, V. Narayanamurti,
J. H. Smet, and W. B. Johnson, and the receipt of high quality MBE
material from V. Umansky.
%%%%%%%%%%%%%%%%%%%%%%%%%%%%%%%%%%%%%%%%%%%%%%%%%%%%%%%%%%%%
% Doing Appendix(ices) - Appendix A & B are shown below    %
%%%%%%%%%%%%%%%%%%%%%%%%%%%%%%%%%%%%%%%%%%%%%%%%%%%%%%%%%%%%
%%%%%%%%%%%%%%%%%%%%%%%%%%%%%%%%%%%%%%%%%%%%%%%%%%%%%%%%%%%%
% Doing references:                                %
%%%%%%%%%%%%%%%%%%%%%%%%%%%%%%%%%%%%%%%%%%%%%%%%%%%%%%%%%%%%


\begin{thebibliography}{35}
%%%%%%%%%%%%%%%%%%%%%%%%%%%%%%%%%%%%%%%%%%%%%%%%%%%%%%%%%%%%
% Command and Example                                %
%                                                          %
% \bibitem{REFERENCE_LABEL} AUTHORS NAMES,                 %
% {\it JOURNAL"S NAMES}{\bf VOLUME NUMBER}, PAGE (YEAR).   %
%                                                          %
% Three examples given below.                            %
%%%%%%%%%%%%%%%%%%%%%%%%%%%%%%%%%%%%%%%%%%%%%%%%%%%%%%%%%%%%
\bibitem{1} B. G. Levi, \textit{Physics Today} \textbf{57} (7), 21 (2004).

\bibitem{2} L. V. Butov et al., \textit{Nature} \textbf{418}, 751 (2002); \textit{Phys. Rev. Lett.} \textbf{92}, 117404
(2004).

\bibitem{3} I. B. Spielman et al., \textit{Phys. Rev. Lett.} \textbf{84}, 5808 (2000); M.
Kellogg et al., \textit{Phys. Rev. Lett.} \textbf{88}, 126804
(2002).

\bibitem{4} R. E. Prange and S. M. Girvin, (eds) \textit{The Quantum Hall
Effect}, 2nd Ed., New York: Springer-Verlag, 1990.

\bibitem{5} R. Fitzgerald, \textit{Phys. Today }\textbf{56}
(4), 24-27 (2003).

\bibitem{6} R. G. Mani et al., \textit{Bull. Am. Phys. Soc.} \textbf{46}, p. 972 (2001).

\bibitem{7} M. Zudov, R. Du, J. Simmons, and J. Reno,
\textit{Phys. Rev. B} \textbf{64}, 201311 (2001).

\bibitem{8} P. D. Ye et al., \textit{Appl. Phys. Lett.
}\textbf{79}, 2193 (2001).

\bibitem{9} R. G. Mani et al., \textit{Nature (London)} \textbf{420}, 646 (2002); \textit{Phys.
Rev. B} \textbf{69}, 193304 (2004); cond-mat/0305507;
cond-mat/0306388; \textit{Phys. Rev. B} \textbf{69}, 161306
(2004);  \textit{Phys. Rev. Lett. }\textbf{92}, 146801 (2004); R.
G. Mani, \textit{Physica E }\textbf{22}, 1 (2004);
cond-mat/0407143.

\bibitem{10} M. Zudov et al., \textit{Phys. Rev. Lett.} \textbf{90}, 046807 (2003); C. L. Yang et al., Phys Rev. Lett. 91, 096803
(2003); R. R. Du et al., \textit{Physica E }\textbf{22}, 7 (2004).

\bibitem{11} S. I. Dorozhkin, \textit{JETP Lett.} \textbf{77}, 577 (2003).

\bibitem{12} S. A. Studenikin et al.,
\textit{Sol. St. Comm.} \textbf{129}, 341 (2004)

\bibitem{13} A. E. Kovalev et al., \textit{Sol. St. Comm.} \textbf{130}, 379
(2004).

\bibitem{14} V. I. Ryzhii, \textit{Sov. Phys. - Sol. St.}
\textbf{11}, 2078-2080 (1970).

\bibitem{15} J. C. Phillips, \textit{Sol. St. Comm. }\textbf{127},
233 (2003).

\bibitem{16} A. Durst et al.,
\textit{Phys. Rev. Lett.} \textbf{91}, 086803 (2003); A. C. Durst
and S. M. Girvin, \textit{Science} \textbf{304}, 1752 (2004).

\bibitem{17} A. V. Andreev, I. L. Aleiner, and A. J. Millis, \textit{Phys. Rev. Lett.} \textbf{91},
056803 (2003).

\bibitem{18} J. Shi and X. C. Xie, \textit{Phys. Rev. Lett.}
\textbf{91}, 086801 (2003).

\bibitem{19} A. A. Koulakov and M. E. Raikh,
\textit{Phys. Rev. B} \textbf{68}, 115324 (2003).

\bibitem{20} F. S. Bergeret, B. Huckestein, and A. F. Volkov,
\textit{Phys. Rev. B} 67, 241303 (2003).

\bibitem{21} I. A. Dmitriev, A. D. Mirlin, and D. G. Polyakov, \textit{Phys. Rev. Lett.}
\textbf{91}, 226802 (2003).

\bibitem{22} X. L. Lei and S. Y. Liu, \textit{Phys. Rev. Lett.} \textbf{91}, 226805 (2003).

\bibitem{23} V. Ryzhii and V. Vyurkov, \textit{Phys. Rev.
B} \textbf{68}, 165406 (2003).

\bibitem{24} D. H. Lee and J. M. Leinaas, \textit{Phys. Rev. B} \textbf{69}, 115336 (2004).

\bibitem{25} V. Ryzhii, \textit{Phys. Rev. B} \textbf{68},
193402 (2003).

\bibitem{26} M. G. Vavilov and I. L. Aleiner,
\textit{Phys. Rev. B} \textbf{69}, 035303 (2004).

\bibitem{27} A. F. Volkov and V. V. Pavlovskii, \textit{Phys. Rev. B }\textbf{69},
125305 (2004).

\bibitem{28} V. Ryzhii and A. Satou, \textit{J. Phys. Soc. Jpn.}
\textbf{72}, 2718 (2003).

\bibitem{29} K. Park, \textit{Phys. Rev. B} \textbf{69}, 201301 (2004).

\bibitem{30} V. Ryzhii, R. Suris, and B. Shchamkhalova \textit{Physica E} \textbf{22}, 13 (2004).

\bibitem{31} I. A. Dmitriev et al., cond-mat/0310668.

\bibitem{32} S. Fujita and K. Ito, cond-mat/0402174.

\bibitem{33} M. Vavilov et al., cond-mat/0405377.

\bibitem{34} C. Joas, M. E. Raikh, and F. von Oppen, cond-mat/0405443.

\bibitem{35} A. E. Patrakov and I. L. Lyapilin, cond-mat/0405517.

\end{thebibliography}
\end{document}